\documentclass[journal,twoside,web]{ieeetran}
\usepackage{cite}
\usepackage{hyperref}

\usepackage{graphicx} % for pdf, bitmapped graphics files
\usepackage{epsfig} % for postscript graphics files
\usepackage{mathptmx} % assumes new font selection scheme installed
\usepackage{times} % assumes new font selection scheme installed
\usepackage{amsmath} % assumes amsmath package installed
\usepackage{amssymb,amsbsy}  % assumes amsmath package installed
\usepackage{comment}
\usepackage{siunitx}
\usepackage[ruled]{algorithm2e}
\usepackage{subcaption}
\usepackage{verbatim}
\usepackage{orcidlink}
\allowdisplaybreaks

\newtheorem{prop}{Proposition} %Proposition environment
\DeclareSIUnit\gcdw{gCDW} %Units
\newcommand\figref[1]{Fig.~\ref{#1}} %Figure reference

\title{\LARGE \bf
Modeling Cell Size Distribution with Heterogeneous Flux Balance Analysis
}

\author{Michiel Busschaert\textsuperscript{\orcidlink{0000-0002-0241-2784}}, Florence H. Vermeire\textsuperscript{\orcidlink{0000-0002-5607-8152}}, and Steffen Waldherr\textsuperscript{\orcidlink{0000-0002-0936-579X}}, \IEEEmembership{Member, IEEE}
\thanks{Manuscript received on March 17 2023. Revised on May 16 2023. Accepted on May 19 2023. This work has been funded by Fonds voor Wetenschappelijk Onderzoek Vlaanderen (FWO), grant number G066621N. \textit{(Corresponding author: Michiel Busschaert.)}}
\thanks{M. Busschaert and F. Vermeire are with the Department of Chemical Engineering,
        KU Leuven, 3001 Leuven, Belgium (e-mail: michiel.busschaert@kuleuven.be and florence.vermeire@kuleuven.be).}
\thanks{S. Waldherr is with the Department of Functional and Evolutionary Ecology, University of Vienna, 1030 Vienna, Austria (e-mail: steffen.waldherr@univie.ac.at).}}

\makeatletter
\def\ps@IEEEtitlepagestyle{%
  \def\@oddfoot{}%
  \def\@oddhead{\mycopyrightnotice}\relax
  \def\@evenhead{\@IEEEheaderstyle\thepage\hfil\leftmark\hbox{}}\relax
  \def\@evenfoot{}%
}
\def\mycopyrightnotice{%
  \begin{minipage}{\textwidth}
  \centering \small
  Copyright~\copyright~2023 IEEE. Personal use of this material is permitted. Permission from IEEE must be obtained for all other uses, in any current or future media, including reprinting/republishing this material for advertising or promotional purposes, creating new collective works, for resale or redistribution to servers or lists, or reuse of any copyrighted component of this work in other works.
  \end{minipage}
}
\makeatother

\begin{document}

\maketitle

%%%%%%%%%%%%%%%%%%%%%%%%%%%%%%%%%%%%%%%%%%%%%%%%%%%%%%%%%%%%%%%%%%%%%%%%%%%%%%%%

\begin{abstract}

For over two decades, Flux Balance Analysis (FBA) has been successfully used for predicting growth rates and intracellular reaction rates in microbiological metabolism. An aspect that is often omitted from this analysis, is segregation or heterogeneity between different cells. In this work, we propose an extended FBA method to model cell size distributions in balanced growth conditions. Hereto, a mathematical description of the concept of balanced growth in terms of cell mass distribution is presented. The cell mass distribution, quantified by the Number Density Function (NDF), is affected by cell growth and cell division. An optimization program is formulated in a general manner in which the NDF, average cell culture growth rate and reaction rates per cell mass are treated as optimization variables. As qualitative proof of concept, the methodology is illustrated on a core carbon model of \textit{Escherichia coli} under aerobic growth conditions. This illustrates feasibility and applications of this method, while indicating some shortcomings intrinsic to the simplified biomass structuring and the time invariant approach.

\end{abstract}

%%%%%%%%%%%%%%%%%%%%%%%%%%%%%%%%%%%%%%%%%%%%%%%%%%%%%%%%%%%%%%%%%%%%%%%%%%%%%%%%

%Keywords obtained from: https://www.ieee.org/content/dam/ieee-org/ieee/web/org/pubs/ieee-thesaurus.pdf

%%%%%%%%%%%%%%%%%%%%%%%%%%%%%%%%%%%%%%%%%%%%%%%%%%%%%%%%%%%%%%%%%%%%%%%%%%%%%%%%

\section{INTRODUCTION}

\IEEEPARstart{H}{eterogeneity}, or physiological and phenotypical diversity, in cell populations arises naturally at different levels. Heterogeneity in protein expression seems to be caused by stochastic processes, either arising from a biological evolutionary mechanism, or as a response to a changing environment \cite{lidstrom2010role,gonzalez2017heterogeneity}. In particular, heterogeneity in cell mass is a direct result of cell growth and cell division, where the interplay between both processes is essential to maintain cell size homeostasis \cite{jun2018fundamental,noris2015why,rhind2021cell,taheri2015cell}. There are several reasons for the cell size control, for example, large cells may experience limited transport via diffusion \cite{marshall2012what}. Simulation-based models of heterogeneity in populations are separated into two classes, individual-based models and population balance models \cite{gonzalez2017heterogeneity}. In the former, the heterogeneity is accounted for by simulating a finite number of cells, whereas in the latter, the population dynamics are described by a density function subject to averaged processes \cite{imig2015individual}.

At the core of cellular processes lies the metabolism, i.e., all different intracellular reactions and metabolites constituting a metabolic network, responsible for consuming one or multiple substrates leading to the production of biomass and by-products. A relatively simply yet popular method to resolve the metabolic network in absence of detailed kinetics, is Flux Balance Analysis (FBA) \cite{varma1994metabolic,orth2010what,heirendt2019creation}. This method estimates fluxes based on the optimization of an objective, often cell growth rate, inspired by evolutionary principles. Although FBA in itself does not consider heterogeneity of cell size, several individual-based models integrating FBA have been reported in literature \cite{damiani2017popfba,labhsetwar2013heterogeneity,tourigny2021simulating,harcombe2014metabolic}.

This paper attempts to include cell mass heterogeneity and cell division described by a population balance model within the FBA framework, in order to model the cell size distribution. To the best of our knowledge, no prior publication describes this concept. The contributions in this paper are 1) formulating a quasi-steady state condition taking cell mass into account, 2) modifying the FBA program to include heterogeneity on the level of individual cell and 3), evaluating the methodology using a metabolic network describing the core carbon pathways in \textit{Escherichia coli} \cite{orth2010reconstruction}.

%%%%%%%%%%%%%%%%%%%%%%%%%%%%%%%%%%%%%%%%%%%%%%%%%%%%%%%%%%%%%%%%%%%%%%%%%%%%%%%%

\section{BACKGROUND INFORMATION}

\subsection{Flux Balance Analysis}\label{ssec:FBA}
A core aspect in FBA is the inclusion of the metabolic network. This network, specific per organism, describes the ensemble of different biochemical reactions involved in the conversion of substrate molecules into biomass and by-products \cite{varma1994metabolic,heirendt2019creation,feist2009reconstruction}. The model scope determines the dimensions of this metabolic network, ranging from tens to thousands of metabolites and reactions. The metabolic network is mathematically represented by the so-called stoichiometric matrix $S$ containing the reaction coefficients per metabolite for each reaction. The dimensions of its rows and columns equal the number of metabolites and reactions, respectively. The biomass growth reaction is typically lumped into a single reaction involving several metabolites, resulting in an unstructured model, where biomass is expressed in gram cell dry weight (\SI{}{\gcdw}).

In FBA, it is assumed that the organism grows under quasi-steady state conditions, often termed balanced growth. Under quasi-steady state conditions, the intensive properties of the culture (e.g. intracellular metabolite concentration) remain constant, while extensive properties (e.g. total biomass) increase. Neglecting dilution due to growth, the quasi-steady state mass balance is formulated as \cite{varma1994metabolic}
\begin{equation}\label{eq:qss_balance}
    S\cdot \mathbf{v} = \mathbf{0}.
\end{equation}
Here $S$ represents stoichiometric matrix, and the vector $\mathbf{v}$ represents the reaction fluxes in the metabolic network (units: $\SI{}{\mole \per\gcdw\per\hour}$ or $\SI{}{\per\hour}$). The specific rate of production of a metabolite in a reaction is given by the product of the reaction flux with its corresponding coefficient in the stoichiometric matrix $S$.

FBA typically assumes a maximization of the growth rate $\mu = \mathbf{c}^\intercal \mathbf{v}$, with $\mathbf{c}$ containing weighting factors to the biomass produced per reaction. Often, $\mathbf{c}$ is $0$ in all but one entry, with the nonzero entry corresponding to the biomass growth reaction. Optimizing the growth rate $\mu$ subject to (s.t.) the quasi-steady state metabolism and reaction flux constraints, the FBA program is given as follows \cite{varma1994metabolic,orth2010what,zomorrodi2016optimization}
\begin{equation}\label{eq:fba}
    \begin{split}
        \max_\mathbf{v} & \quad \mu = \mathbf{c}^\intercal \mathbf{v}, \\
        \text{s.t.} & \quad S\cdot \mathbf{v} = \mathbf{0}, \\
        & \quad \mathbf{v_\text{lb}} \leq \mathbf{v} \leq \mathbf{v_\text{ub}}.
    \end{split}
\end{equation}
The vectors $\mathbf{v_\text{lb}}$ and $\mathbf{v_\text{ub}}$ impose bounds on reaction fluxes, defined by thermodynamic (e.g. reaction irreversibility) and kinetic constraints (e.g. maximum uptake rate). The FBA program in \eqref{eq:fba} is a linear optimization program, which can be solved efficiently \cite{zomorrodi2016optimization} with standard solvers such as ILOG CPLEX (IBM).

The objective, i.e. maximization of growth rate, follows from an evolutionary argument. Over a large timespan, natural occurring species are assumed to have evolved to utilize resources close to optimality. It should be noted that a variety of objective functions for \eqref{eq:fba} have been formulated \cite{zomorrodi2016optimization,schuetz2007systematic}, either describing a similar evolutionary assumption (e.g. ATP yield maximization) or relying on different considerations. As result of the optimization, one can \textit{in silico} predict the reaction fluxes $\mathbf{v}$ of the organism under changes in growth conditions or under genetic modification \cite{orth2010what}.

\subsection{Population Balance Modeling}\label{ssec:PBM}
Population systems are encountered in many applications in engineering, including cellular systems. The population is described as a collection of individual particles, each characterized with one or multiple states, that undergo the same type of processes. Population Balance Models (PBM) provide a deterministic approach to describe the time evolution of such population systems with heterogeneity among its individuals \cite{ramkrishna2000population,ramkrishna2014population}. In cell populations, this heterogeneity, the particle state, may be e.g. cell mass or size, cell age or enzyme content \cite{mantzaris2004cell}.

In this work, the cell mass is considered as single state $x$, that is expected to evolve over time described by $\dot{x} = \mu(x)x$. The biomass distribution of cells is represented by the Number Density Function $n(t,x)$. As implied by the name, the NDF $n(t,x)$ denotes an (unnormalized) probability density function, such that the total number of particles with state $x$ belonging to a domain $\left[x_1,x_2\right]$ is equal to $\int_{x_1}^{x_2} n(t,\xi)d\xi$.

The time evolution of the NDF $n(t,x)$ for a cell culture undergoing cell growth and division is given by the following Population Balance Model:
\begin{equation}\label{eq:pbm_general}
    \begin{split}
        \frac{\partial n(t,x)}{\partial t} & + \frac{\partial \mu(x)xn(t,x)}{\partial x} = \\ & -\gamma(x)n(t,x)+\int_x^\infty \beta(x,x')\gamma(x')n(t,x')dx'.
    \end{split}
\end{equation}
On the left-hand side, the first term represents a rate of change over time, and the second term corresponds to cell growth. The right-hand side represents cell division at rate $\gamma(x)$ and division kernel $\beta(x,x')$. The negative term indicates dividing mother cells, whereas the positive term indicates divided daughter cells from larger cells. Equation \eqref{eq:pbm_general} is derived based on a conservation of number of particles. For further information on the origin of \eqref{eq:pbm_general}, see e.g. \cite{ramkrishna2000population}.

Typically the cell population interacts with a continuous phase, e.g. concentrations of nutrients, which in turn may influence kinetic rates and add additional differential equations to the overall model \eqref{eq:pbm_general} \cite{mantzaris2004cell}. Furthermore, standard no-flux boundary conditions are defined \cite{ramkrishna2000population,mantzaris2004cell}, meaning cells do not leave/enter the domain $[0,\infty)$,
\begin{equation}
    \mu(x)xn(t,x) \rightarrow 0, \text{ for } x\rightarrow 0\text{ or }x\rightarrow \infty.
\end{equation}
The cell division rate $\gamma(x)$ indicates how frequently cells in the population divide. The division rate $\gamma(x)$ is assumed to be expressed solely as a function of cell mass $x$. Modeling the division rate remains an open issue in literature, and other dependencies, e.g. cell age or increased size, have been considered in other works, see e.g. \cite{jun2018fundamental,taheri2015cell,totis2021population}

%Modeling the division rate still remains an open issue in literature. In one perspective, three cellular size control mechanisms are proposed, being the `sizer', `adder' and `timer' mechanisms \cite{taheri2015cell,sauls2015adder,jun2018fundamental,totis2021population}. In short, the `sizer' mechanism suggests that cells divide upon reaching a critical size. The `adder' mechanism supports the idea that cells divide once the cell mass has increased with a critical amount (relative to size at birth). Finally the `timer' claims that cells divide at a certain critical age (this is, elapsed time since birth). By modeling the division rate with $\gamma(x)$ as in \eqref{eq:pbm_general}, a `sizer' mechanism is explicitly assumed. The `adder' and `timer' mechanisms can be modeled explicitly via different formulations of the PBM \cite{taheri2015cell,jun2018fundamental}. As \cite{totis2021population} demonstrates, the apparent behavior of a PBM as in \eqref{eq:pbm_general} can still appear `adder' or `timer'-like.

The division kernel $\beta(x,x')$ describes the average number of daughter cells with cell mass $x$ that are born after division of a mother cell with mass $x'$, such that $\beta(x,x') = 0$ for $x > x'$. In addition the kernel must satisfy mass conservation, this is, $x' = \int_0^{x'} \xi\beta(\xi,x')d\xi$ \cite{ramkrishna2000population,mantzaris2004cell}. Binary division is modeled by a particular kernel $\beta(x,x') = 2\delta(x-x'/2)$, in which a mother cell divides into two equally sized daughter cells. This phenomenon is encountered in \textit{Escherichia coli} and other prokaryotic bacteria \cite{taheri2015cell,chien2012cell}.

The total biomass can be expressed as
\begin{equation}\label{eq:biomass}
    B(t) = \int_0^\infty \xi n(t,\xi)d\xi.
\end{equation}
Multiplying the PDE in \eqref{eq:pbm_general} with cell mass $x$ and integrating over the cell mass domain $[0,\infty)$ it is possible to show the time evolution of the total biomass to be given as
\begin{equation}\label{eq:biomass_derivative}
    \frac{dB(t)}{dt} = \int_0^\infty \mu(\xi)\xi n(t,\xi)d\xi.
\end{equation}
Thus the total biomass change depends explicitly only on the specific growth rate $\mu(x)$, but it can depend implicitly on cell division.

%%%%%%%%%%%%%%%%%%%%%%%%%%%%%%%%%%%%%%%%%%%%%%%%%%%%%%%%%%%%%%%%%%%%%%%%%%%%%%%%

\section{HETEROGENEOUS FBA}\label{sec:heterogeneous_FBA}
In the following Subsection \ref{ssec:coupled_dynamics}, Equation \eqref{eq:pbm_general} is reformulated to include the metabolite dynamics. In Subsection \ref{ssec:qss_conditions}, the quasi-steady state conditions for the PBM are formulated. In Subsection \ref{ssec:hfba_description}, the result hereof is combined with the assumption of maximizing growth rate, to extend standard FBA with cell size heterogeneity, which is denoted as heterogeneous FBA.

\subsection{Coupling the PBM with metabolite dynamics}\label{ssec:coupled_dynamics}
Equation \eqref{eq:pbm_general} describes the evolution of a cell population. The metabolite dynamics are included based on the metabolic network description. This coupling is important to account for the changes in metabolite concentrations over time and results in the set of equations,
\begin{equation}\label{eq:pbm_cell}
    \begin{split}
        & \frac{\partial n(t,x)}{\partial t} + \frac{\partial (\mu(x,\mathbf{X},\mathbf{Y})xn(t,x))}{\partial x} = \\ & \qquad -\gamma(x)n(t,x) + \int_x^\infty \beta(x,x')\gamma(x')n(t,x')dx', \\
        & \frac{d}{dt} \begin{pmatrix} \mathbf{X}(t) \\ \mathbf{Y}(t) \end{pmatrix} = \int_0^\infty \begin{pmatrix} S_X \\ S_Y \end{pmatrix}\cdot \mathbf{v}(\xi,\mathbf{X},\mathbf{Y})\xi n(t,\xi)d\xi.
    \end{split}
\end{equation}
Here, $\mathbf{X}(t)$ and $\mathbf{Y}(t)$ represents the amount of intracellular and extracellular metabolites, respectively. The matrices $S_X$ and $S_Y$ represent submatrices of the stoichiometric matrix $S$. The integral term in the metabolite dynamics sums the production or consumption from each individual cell mass interval. In practice, the growth rate $\mu(x,\mathbf{X},\mathbf{Y})$ and reaction fluxes $\mathbf{v}(x,\mathbf{X},\mathbf{Y})$ depend on metabolite concentrations (e.g. Michaelis-Menten kinetics). These rate expressions as functions of $\mathbf{X}(t)$ and $\mathbf{Y}(t)$ are often unknown, such that \eqref{eq:pbm_cell} cannot be integrated directly.

\subsection{Quasi-steady state conditions}\label{ssec:qss_conditions}
The PBM in \eqref{eq:pbm_cell} describes a time-dependent evolution of the NDF. One is often interested in stationary behavior, corresponding to so-called balanced growth or quasi-steady state conditions. Here, the total biomass increases exponentially, while the relative composition remains constant. This is a base concept in FBA \cite{varma1994metabolic}, as well as in Resource Balance Analysis (RBA) \cite{goelzer2011cell}, in which biomass is modeled as a structured composition. Describing growth under quasi-steady state has the advantage to reduce computational effort while still describing a relevant process. A disadvantage is that time-dependent process cannot be described.

In quasi-steady state conditions, metabolite concentrations remain constant. This implies that during balanced growth, the reaction fluxes $\mathbf{v}(x,\mathbf{X},\mathbf{Y})$ and growth rate $\mu(x,\mathbf{X},\mathbf{Y})$ remain constant in $\mathbf{X}$ and $\mathbf{Y}$. For the remainder of the text, the unknown dependencies on $\mathbf{X}$ and $\mathbf{Y}$ for these reaction fluxes and growth rate are omitted in notation of the reaction flux vector, i.e. $\mathbf{v}(x)$ and $\mu(x)$, and are instead to be resolved based on optimization under constraints, as in standard FBA.

To extend the concept of balanced growth with cell mass heterogeneity, it is assumed that the ratio of the biomass of cells with mass in the interval $[x,x+\Delta x]$ to the total biomass remains constant. Mathematically, this is
\begin{equation}
    \frac{d}{d t} \left(\frac{\int_x^{x+\Delta x} \xi n(t,\xi) d\xi}{\int_0^\infty \xi n(t,\xi) d\xi} \right) = 0.
\end{equation}
The equation above can be rewritten as
\begin{equation}
    \begin{split}
        \int_x^{x+\Delta x} & \frac{\partial \xi n(t,\xi)}{\partial t} d\xi = \\
        & \frac{\int_x^{x+\Delta x} \xi n(t,\xi) d\xi}{\int_0^\infty \xi n(t,\xi) d\xi}\frac{d}{d t}\int_0^\infty \xi n(t,\xi) d\xi. \\
    \end{split}
\end{equation}
For infinitesimal small intervals, i.e. $\Delta x \rightarrow 0$, and dividing both sides by the factor $x\Delta x$, this reduces to
\begin{equation}\label{eq:qss_condition}
    \frac{\partial n(t,x)}{\partial t} = \frac{n(t,x)}{\int_0^\infty \xi n(t,\xi)d\xi}\int_0^\infty \mu(\xi)\xi n(t,\xi)d\xi,
\end{equation}
using \eqref{eq:biomass} and \eqref{eq:biomass_derivative}. The coefficient to $n(t,x)$ on the right-hand side of \eqref{eq:qss_condition} can be interpreted as the average growth, weighted against the amount of biomass per individual cell mass $x$. The average growth rate $\bar{\mu}$ is defined as
\begin{equation}
    \bar{\mu} = \frac{1}{\int_0^\infty \xi n(t,\xi) d\xi}\int_0^\infty \mu(\xi)\xi n(t,\xi)d\xi.
\end{equation}
This average growth rate corresponds to the specific growth rate of the entire culture. As a consequence, by substituting \eqref{eq:qss_condition} into the PBM \eqref{eq:pbm_general}, the time-dependency of $n(t,x)$ describes an exponential function with $\bar{\mu}$, with the cell mass-dependency described by an integro-differential equation. One may regard this as the cell growth and cell division being perfectly balanced out against each other, such that the size distribution, i.e. cell mass heterogeneity, remains constant. This formulation is consistent with other time-invariant solutions to cell population models \cite{imig2015individual}.

\subsection{Heterogeneous FBA}\label{ssec:hfba_description}
The quasi-steady state conditions, as derived in previous subsection, are formulated as part of a wider constraint-based optimization framework. Similar to the reasoning in FBA (Subsection \ref{ssec:FBA}), the unknown reaction fluxes $\mathbf{v}(x)$ and growth rate $\mu(x)$ are resolved assuming optimization of an objective function, here corresponding to the maximization of total biomass growth $\bar{\mu}B$, where $B$ is the total biomass. As the total biomass $B$ increases exponentially, the average specific growth rate $\bar{\mu}$ is chosen as objective. Due to the heterogeneous cell mass considered, the maximization of growth rate is considered over the total population, rather than individual cell masses.

The growth rate $\mu(x)$ can be substituted as $\mu(x) = \mathbf{c}(x)^\intercal \mathbf{v}(x)$. Here, $\mathbf{c}(x)$ contains the coefficients to determine the mass-dependent growth rate $\mu(x)$, which can be used to describe the biomass production reaction changing with individual cell mass, as the cell composition differs between cells of different mass. Applying quasi-steady state conditions \eqref{eq:qss_condition} to the PBM with continuous dynamics \eqref{eq:pbm_cell}, the following optimization program is proposed to describe FBA with cell mass heterogeneity:
\begin{subequations}\label{eq:hfba}
    \begin{align}
        \max_{n(x),\mathbf{v}(x),\bar{\mu}} & \bar{\mu},\label{seq:hfba_a} \\
        \text{s.t. } & \bar{\mu}n(x) \leq - \frac{d}{dx}\left(\mathbf{c}(x)^\intercal\mathbf{v}(x)xn(x)\right) \label{seq:hfba_b} \\
        & \quad -\gamma(x)n(x)+\int_x^\infty \beta(x,x')\gamma(x')n(x')dx',\nonumber \\
        & \mathbf{c}(x)^\intercal\mathbf{v}(x)xn(x) \rightarrow 0,\text{ for } x\rightarrow 0 \text{ or } x\rightarrow\infty, \label{seq:hfba_c} \\
        & S_X\cdot \mathbf{v}(x)xn(x) = \mathbf{0}, \label{seq:hfba_d} \\
        & \int_0^\infty S_Y\cdot \mathbf{v}(\xi)\xi n(\xi)d\xi \geq \mathbf{b}, \label{seq:hfba_e} \\
        & \int_0^\infty \mathbf{c}(\xi)^\intercal \mathbf{v}(\xi)\xi n(\xi)d\xi \geq \bar{\mu}B, \label{seq:hfba_f} \\
        & \int_0^\infty \xi n(\xi)d\xi = B, \label{seq:hfba_g} \\
        & \mathbf{v}_\text{lb}(x)xn(x) \leq \mathbf{v}(x)xn(x) \leq \mathbf{v}_\text{ub}(x)xn(x), \label{seq:hfba_h} \\
        & 0 \leq n(x). \label{seq:hfba_i}
    \end{align}
\end{subequations}
Here, the objective as optimization of average growth rate $\bar{\mu}$ is given in \eqref{seq:hfba_a}. Equation \eqref{seq:hfba_b} tracks the balance between growth and division to sustain balanced growth, along with the boundary conditions \eqref{seq:hfba_c}. Whereas formulations of the PBM typically define the population balance as an equality, a nonstrict inequality is used in \eqref{seq:hfba_b}. The motivation behind this change is discussed at the end of this section. Constraint \eqref{seq:hfba_d} denotes a mass balance of intracellular metabolites in quasi-steady state at every cell mass interval, similar to FBA. Constraint \eqref{seq:hfba_e} provides a limit on total uptake $\mathbf{b}$ of extracellular components\footnote{By convention, a negative exchange reaction flux denotes uptake \cite{zomorrodi2016optimization}.}. Constraint \eqref{seq:hfba_f} denotes a mass balance for the total biomass, based on the average specific growth rate $\bar{\mu}$. Similar to \eqref{seq:hfba_b}, the equality constraint is replaced as an inequality constraints. The total biomass $B$ is fixed in Constraint \eqref{seq:hfba_g}. This constraint is required, as it provides a sort of normalization to the NDF. Alternatively, this constraint could be formulated as a normalization of $n(x)$, in which case the total biomass $B$ is not defined \textit{a priori}. Constraint \eqref{seq:hfba_h} provides lower and upper bounds on reaction fluxes, $\mathbf{v}_\text{lb}(x)$ and $\mathbf{v}_\text{ub}(x)$, that are used to impose reaction irreversibility or to include information on maximum/minimum kinetic rates. Constraint \eqref{seq:hfba_i} ensures that the NDF $n(x)$ is positive.

The optimization problem in \eqref{eq:hfba} is discretized using a finite volume scheme with first order upwind discretization \cite{motz2002comparison} to resolve the derivative to $x$ in \eqref{seq:hfba_b}. The optimization problem in \eqref{eq:hfba}, after discretization, is nonlinear. However, note that \eqref{eq:hfba} is expressed in such way that the term $\mathbf{v}(x)$ only appears in terms of $\mathbf{v}(x)xn(x)$. By defining $n(x)$ and $\mathbf{v}(x)xn(x)$ as the optimization variables, for a fixed value of $\bar{\mu}$, the constraints in the discretized problem are linear in the optimization variables. In this case, the value of $\bar{\mu}$ determines whether the problem is feasible or infeasible, and the problem can be solved as a linear feasibility program, similar to the approach followed in \cite{goelzer2011cell,goelzer2011bacterial}, by following proposition:
\begin{prop}\label{prop:mu_opt}
    If the optimization program \eqref{eq:hfba} has a bounded solution (i.e. is infeasible for $\bar{\mu}\rightarrow\infty$), then there exists an optimal $\bar{\mu}^*$ such that \eqref{eq:hfba} is feasible for all $\bar{\mu} \leq \bar{\mu}^*$, and infeasible for any $\bar{\mu} > \bar{\mu}^*$.
\end{prop}
\begin{IEEEproof}
    The feasible subspace for \eqref{eq:hfba} for a certain $\bar{\mu}_1$ is a subspace of the feasible subspace for any $\bar{\mu}_2 \leq \bar{\mu}_1$. This can easily be verified from Constraints \eqref{seq:hfba_b} and \eqref{seq:hfba_f} being formulated as inequalities. In addition, a maximum $\bar{\mu}^*$ must exist and be finite, as the solution must be bounded, as \eqref{eq:hfba} is infeasible for $\bar{\mu}\rightarrow\infty$.
\end{IEEEproof}
This proposition shows that the problem \eqref{eq:hfba} may be solved as a linear feasibility program, returning the maximum $\bar{\mu}^*$. Although Constraints \eqref{seq:hfba_b} and \eqref{seq:hfba_f} would be expected as equalities from a theoretical point of view, these are written as inequalities, in part to allow us to prove this proposition. The use of inequality signs in \eqref{seq:hfba_b} and \eqref{seq:hfba_f} may be regarded as a capacity requirement, i.e. sufficient growth and division occurs over all cell masses $x$ to sustain balanced growth at a fixed average growth rate $\bar{\mu}$. At any average growth rate larger than $\bar{\mu}^*$, there is insufficient capacity to maintain the quasi-steady state condition over a range of cell masses. From numerical simulations on two small-scale metabolic networks \cite{orth2010reconstruction,waldherr2015dynamic}, it is observed that solving \eqref{eq:hfba} with \eqref{seq:hfba_b} and \eqref{seq:hfba_f} as either equalities or inequalities results in a nearly identical optimal growth rate $\bar{\mu}^*$, with an error that is either zero or orders of magnitude smaller than the optimum. This suggests that both problems have the same optimal solution up to numerical error, however, we could not formally prove this to hold in general.

%%%%%%%%%%%%%%%%%%%%%%%%%%%%%%%%%%%%%%%%%%%%%%%%%%%%%%%%%%%%%%%%%%%%%%%%%%%%%%%%

\section{CASE STUDY ON \textit{E. COLI} CORE CARBON PATHWAY}
To illustrate the performance of the proposed heterogeneous FBA approach, a case study on the model organism \textit{Eschericia coli} is considered. The case study is performed on a small-scale model of \textit{E. coli} metabolism for aerobic growth on glucose. The original model \cite{orth2010reconstruction} contains $72$ different metabolites and $95$ biochemical reactions. Included in these are $20$ extracellular metabolites (and exchange reactions). Two growth conditions are considered, one in which oxygen limits growth, and one where oxygen is a nonlimiting component. The biomass production reaction is constructed based on $16$ precursor metabolites, with $7$ other metabolites produced as by-products. Among these metabolites are energy-related components such as ATP/ADP, to account for, among others, growth related maintenance. 

Three constraints are imposed. First, as in the original model \cite{orth2010reconstruction}, a non-growth-associated ATP maintenance reaction is present, with fixed reaction flux of \SI{8.39}{\milli\mole\per\gcdw\per\hour}. Second, a maximum bound on glucose uptake rate is fixed to \SI{10}{\milli\mole\per\gcdw\per\hour} for all cell masses (as in Constraint \eqref{seq:hfba_h}). Third, in case oxygen is limiting to the growth, a bound on the total oxygen uptake is set to \SI{12}{\milli\mole\per\hour} (as in Constraint \eqref{seq:hfba_e}). A total biomass of $B = \SI{1}{\gcdw}$ is specified. As individual cell mass is on the order of magnitude of several picograms, the cell mass is scaled for numerical reasons.

Additionally, the division rate $\gamma(x)$ and kernel $\beta(x,x')$ are specified. The division kernel is defined as binary division for \textit{E. coli} \cite{taheri2015cell,osella2014concerted}, this is $\beta(x,x') = 2\delta(x-x'/2)$. The division rate $\gamma(x)$ is taken from literature \cite{osella2014concerted}, using the Hill function,
\begin{equation}\label{eq:division_hill}
    \gamma(L(x)) = k\frac{L(x)^m}{h^m+L(x)^m},
\end{equation}
with the parameter values $m = 12$, $h = \SI{5.65}{\micro\meter}$ and $k = \SI{9.3}{\per\hour}$ \cite{osella2014concerted}. The variable $L$ represents the cell size and is directly related to the cell mass $x$. Given that \textit{E. coli} is rod-shaped, this shape is approximated with a cylinder,
\begin{equation}\label{eq:mass_length}
    x(L) = \rho_\textit{E. coli}\frac{\pi D_\text{rod}^2}{4}L.
\end{equation}
An average cell density $\rho_\textit{E. coli}=\SI{1.105}{\pico\gcdw \per\micro\meter\cubed}$ \cite{martinez1981relationship} and average diameter $D_\text{rod} = \SI{1}{\micro\meter}$ \cite{taheri2015cell,marshall2012what} are used. These values remain more or less constant, even upon cell division. Alternative division rates have been reported in literature \cite{taheri2015cell,robert2014division}.

The optimization program \eqref{eq:hfba} is solved over a grid with cell lengths $L$ ranging from \SI{0}{} to \SI{10}{\micro\meter}. This range is discretized into $20$ equidistant subintervals. ILOG CPLEX (IBM) is used to solve linear programs. The optimal average growth rate $\bar{\mu}^*$ is obtained iteratively. The maximum average growth rate $\bar{\mu}^*$ is unique (from Proposition \ref{prop:mu_opt}), although multiple solutions for $n(x)$ and $\mathbf{v}(x)$ may exist. To quantify bounds on the range of optima, a procedure similar to Flux Variability Analysis (FVA) \cite{zomorrodi2016optimization,mahadevan2003effects} is implemented. For different fixed values $x_i$, the individual optimization variables corresponding to the NDF $n(x_i)$ and each reaction flux $v(x_i)$ are taken as objective function separately subject to the constraints \eqref{seq:hfba_b} to \eqref{seq:hfba_i}, with $\bar{\mu} = \bar{\mu}^*$. Maximization and minimization for each objective variable return upper and lower bounds, respectively.

In the case study, the biomass equation and reaction flux bounds are chosen to be modeled constant over cell masses (in more detailed models, this biomass reaction would differ per cell mass). This means that the average fluxes determined in heterogeneous FBA should (approximately) be equal to those of standard FBA \eqref{eq:fba} simulated under the same conditions, which is observed. Note that standard FBA does not calculate the cell size distribution.

\begin{figure*}
    \centering
    \medskip
    \begin{subfigure}[b]{.32\linewidth}
        \includegraphics[width=\linewidth]{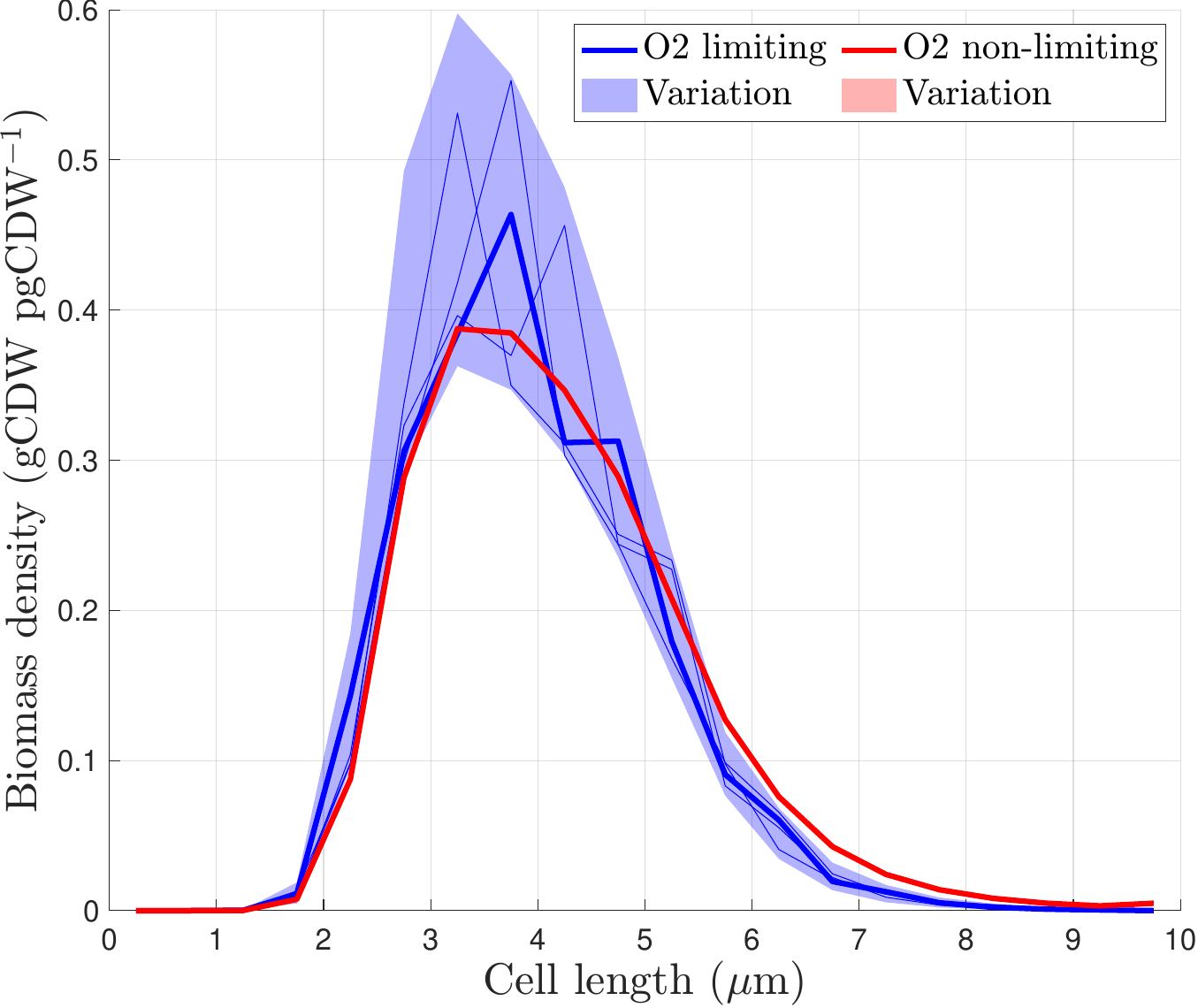}
        \caption{Biomass density function $xn(x)$.} \label{sfig:biomass}
    \end{subfigure}
    \hfill
    \begin{subfigure}[b]{.32\linewidth}
        \includegraphics[width=\linewidth]{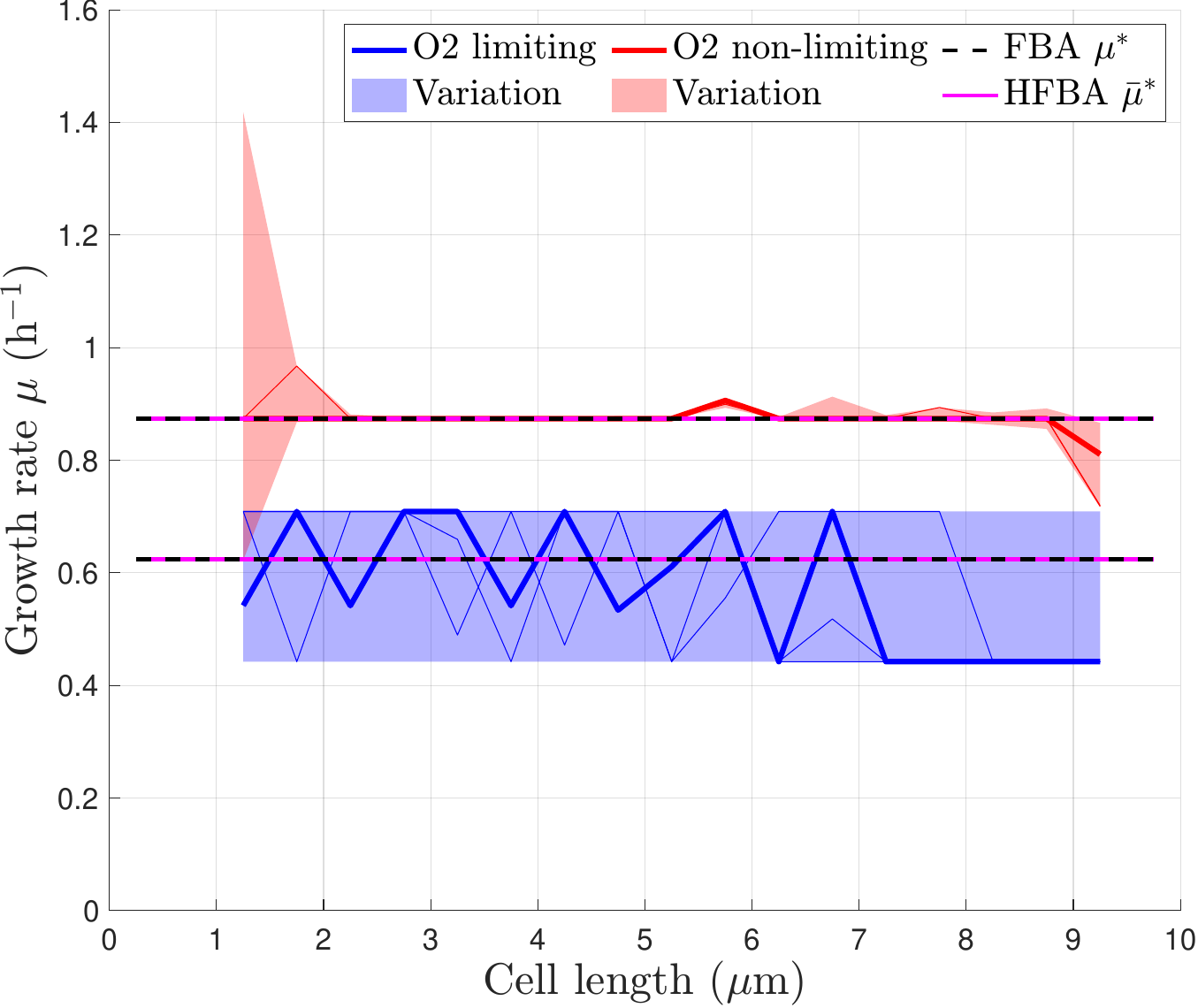}
        \caption{Specific growth rate $\mathbf{c}(x)^\intercal\mathbf{v}(x)$.} \label{sfig:growth}
    \end{subfigure}
    \hfill
    \begin{subfigure}[b]{.32\linewidth}
        \includegraphics[width=\linewidth]{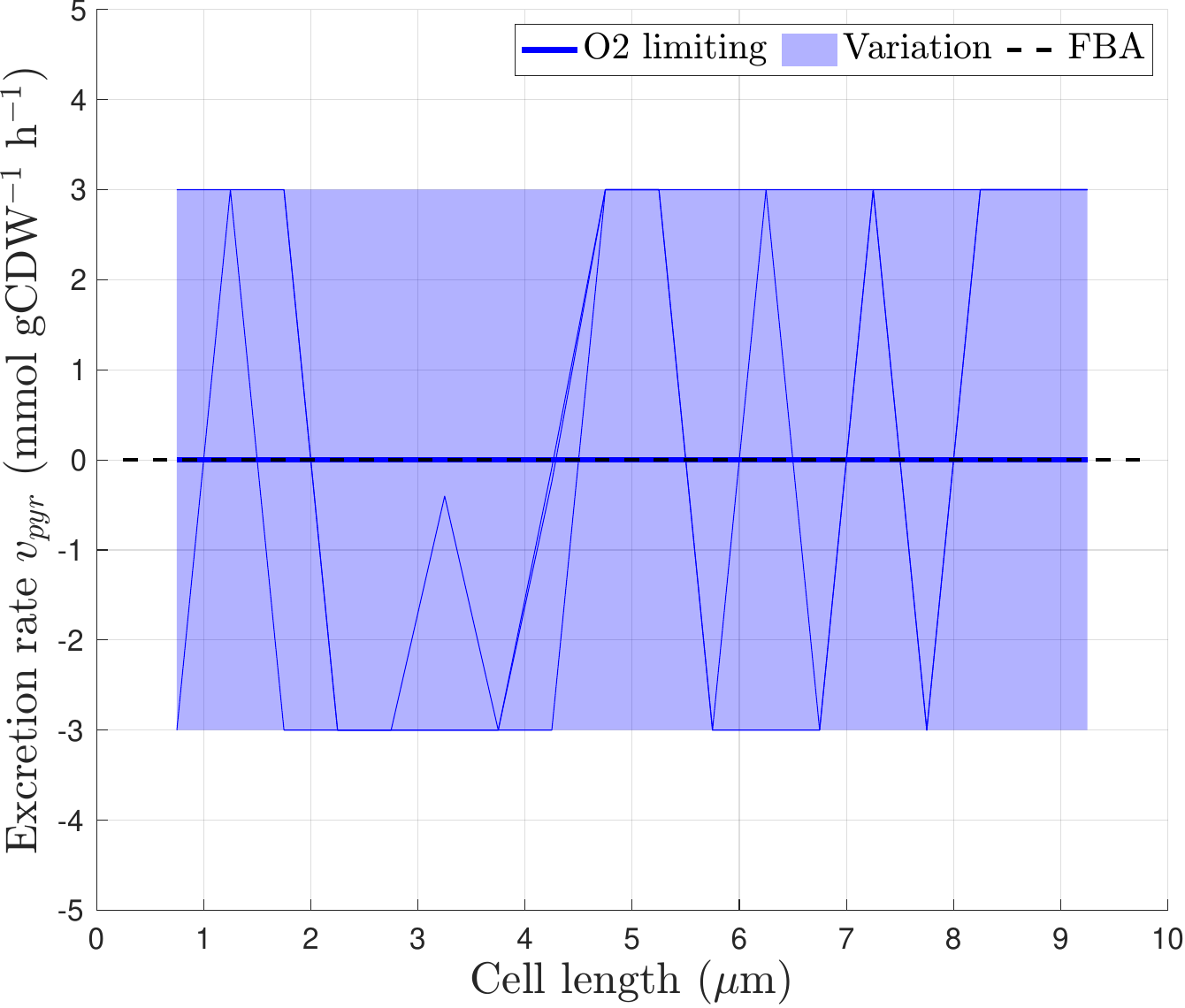}
        \caption{Pyruvate exchange flux $v_\text{pyr}(x)$.}  \label{sfig:exchange}
    \end{subfigure}
    \caption{Simulation result from \eqref{eq:hfba} under conditions with limited oxygen availability (blue, $\bar{\mu}^* = \SI{0.62}{\per\hour}$) and with sufficient oxygen availability (red, $\bar{\mu}^* = \SI{0.87}{\per\hour}$). The shaded region indicates lower and upper bound per cell size interval from a variability analysis, indicating the existence of multiple optimal solutions. The thin lines indicate random alternative optimal solutions. The solution in thick lines are generated without cross-feeding between cells. The black dashed line corresponds to the solution obtained from FBA. In \figref{sfig:exchange}, the results for the nonlimiting oxygen case are omitted for clarity. \figref{sfig:biomass} cannot be obtained with standard FBA.}
    \label{fig:hfba_solution}
\end{figure*}

\figref{fig:hfba_solution} shows results from the heterogeneous FBA program under oxygen limited (blue) and unlimited (red) growth conditions. The optimal average growth rates $\bar{\mu}^* = \SI{0.62}{\per\hour}$ and $\bar{\mu}^* = \SI{0.87}{\per\hour}$ are calculated, which is equal to the solution obtained from standard FBA under the corresponding conditions. As seen from the variability analysis (shaded regions in \figref{fig:hfba_solution}), the feasible subspace at the optimal average growth rate $\bar{\mu}^*$ permits multiple solutions The biomass density distribution is given in \figref{sfig:biomass}, the case of limited oxygen availability (blue) returns a broader solution space compared to the case of no oxygen limitation (red, variability region in biomass density not distinguishable). This effect is caused by the variability in growth rate $\mu(x)$ per cell size, \figref{sfig:growth}. In the case of growth under nonlimiting oxygen conditions, a slight variability is predicted, which is likely caused by numerical errors. In the oxygen limiting case, there is competition regarding oxygen uptake, resulting in some cells masses behaving differently than other cells. On average, the growth rate is equal to the optimal average specific growth rate $\bar{\mu}^*$.

Interestingly, growth strategies under the same optimal growth rate $\bar{\mu}^*$ are predicted, in which cells of certain cell size provide resources (i.e. intracellular metabolites) to cells of different size via exchange reactions, resulting in a more uneven growth distribution between cell sizes. This is demonstrated in \figref{sfig:exchange}, describing the exchange rate of pyruvate, an intermediate metabolite which is part of the biomass production equation. Although in this case study, this effect is caused by a lack of additional `cost' to the exchange reactions, similar \textit{cross-feeding} behavior has been observed experimentally in subpopulations in e.g. yeast \cite{kamrad2023metabolic}. To control these solutions, additional constraints on exchange reactions are applied, as included for the thick lines in \figref{fig:hfba_solution}, or alternatively, an additional quadratic objective may be imposed to minimize the squared sum of exchange reactions.

As a last remark, we return to the motivation of defining Constraints \eqref{seq:hfba_b} and \eqref{seq:hfba_f} as inequalities (Subsection \ref{ssec:hfba_description}). Solving the optimization program with \eqref{seq:hfba_b} and \eqref{seq:hfba_f} expressed as equality constraints, the optimal average growth rate differs with \SI{1.25e-5}{\per \hour} from the optimum obtained with \eqref{seq:hfba_b} and \eqref{seq:hfba_f} as inequalities. This illustrates that switching \eqref{seq:hfba_b} and \eqref{seq:hfba_f} to inequalities is justified, at least in this case study, and it allows for the optimization problem to be solved as a linear feasibility program.

%%%%%%%%%%%%%%%%%%%%%%%%%%%%%%%%%%%%%%%%%%%%%%%%%%%%%%%%%%%%%%%%%%%%%%%%%%%%%%%%

\section{SUMMARY \& CONCLUSION}
This work proposes an extension to Flux Balance Analysis, in which cell size distribution is predicted. Hereto, the framework of Population Balance Models is employed, in combination with a condition on balanced growth on the population level. The problem can be solved numerically efficient as a linear feasibility program. The method is illustrated on a small-scale \textit{E. coli} metabolic network. This illustrates that the program performs well in predicting an optimal average growth rate, but some difficulties may be encountered with respect to the existence of multiple solutions. This can be attributed to the nondetailed representation of the total biomass and reaction kinetics, and may be a matter of choice in imposing reaction flux bounds. As with standard FBA, the growth environment and physiological conditions of the specific organism may affect the suitability of the maximization of specific average growth rate, which should be validated experimentally case by case.

The case study shows that heterogeneous FBA \eqref{eq:hfba} can return identical results to standard FBA \eqref{eq:fba} if modeled under corresponding conditions, which is a minimum requirement to show validity of the definition in \eqref{eq:hfba}, since standard FBA completely \textit{averages out} heterogeneity within the cell culture. Though not demonstrated in the case study, the introduced heterogeneous FBA program facilitates the description of cell composition and kinetic rates per cell mass (via choice of $\mathbf{c}(x)$, $\mathbf{v}_\text{lb}(x)$ and $\mathbf{v}_\text{ub}(x)$). This may allow for the metabolic description of processes such as cross-feeding and cell division by considering a total population growth maximization.

%%%%%%%%%%%%%%%%%%%%%%%%%%%%%%%%%%%%%%%%%%%%%%%%%%%%%%%%%%%%%%%%%%%%%%%%%%%%%%%%

\section*{CODE ACCESS}
Our code and info regarding discretization can be found online on \url{https://github.com/MichielBusschaert/Heterogeneous_FBA}.

%%%%%%%%%%%%%%%%%%%%%%%%%%%%%%%%%%%%%%%%%%%%%%%%%%%%%%%%%%%%%%%%%%%%%%%%%%%%%%%%

\bibliographystyle{IEEEtran.bst}
\bibliography{Bibliography/biblio}

\end{document}